\documentclass[aps,prb,twocolumn,showpacs,showkeys,floatfix]{revtex4}

\usepackage{graphicx,color}

\graphicspath{{figs/}}
\begin{document}
\title{A Fast Uniaxial Compression of the Single-Layer MoS$_{2}$}
\author{Jin-Wu Jiang}
    \altaffiliation{Email address: jwjiang5918@hotmail.com}
    \affiliation{Shanghai Institute of Applied Mathematics and Mechanics, Shanghai Key Laboratory of Mechanics in Energy Engineering, Shanghai University, Shanghai 200072, People's Republic of China}

\date{\today}
\begin{abstract}
The Euler buckling theorem states that the buckling critical strain is an inverse square function of the length for a thin plate in the static compression process. However, the suitability of this theorem in the dynamical process is unclear, so we perform molecular dynamics simulations to examine the applicability of the Euler buckling theorem in case of a fast compression of the single-layer MoS$_{2}$. We find that the Euler buckling theorem is not applicable in such dynamical process, as the buckling critical strain becomes a length-independent constant in the buckled system with many ripples. However, the Euler buckling theorem can be resumed in this dynamical process after restricting the theorem to an individual ripple in the buckled structure.

\end{abstract}

\maketitle
\pagebreak

Molybdenum Disulphide (MoS$_{2}$) has attracted considerable attention in recent years on its electronic, thermal, or mechanical properties,\cite{KamKK,WangQH2012nn,ChhowallaM,gomezAM2012,Castellanos-GomezA2012nrl,HuangW,BertolazziS,CooperRC2013prb1,CooperRC2013prb2,VarshneyV,LiuX2013apl,Castellanos-GomezA2013nl,JiangJW2013mos2,JiangJW2013sw,JiangJW2013bend,JiangJW2013mos2resonator} as it is one of the useful two-dimensional materials. Different two-dimensional materials (eg. graphene and MoS$_{2}$) have complementary physical properties. Therefore, experimentalists have combined graphene and MoS2 in specific ways to create heterostructures that mitigate the negative properties of each individual constituent.\cite{BritnellL2013sci} However, the temperature change will lead to some mechanical compression/tension on the heterostructure, because of different thermal expansion coefficient of graphene and MoS$_{2}$.\cite{BaoW2009nn} This thermal-induced mechanical compression will intrigue the buckling of some layers in the sandwich structure, as the buckling critical strain is usually very low for layered materials. Hence, it is important to investigate the buckling phenomenon, including the single-layer MoS$_{2}$ (SLMoS$_{2}$).

The buckling critical strain of a thin plate can be captured by the Euler buckling theorem, i.e., $\epsilon_{c}  =  -\frac{4\pi^{2}D}{C_{11}L^{2}}$, where $D$ is the bending modulus and $C_{11}$ is the in-plane tension stiffness. $L$ is the length of the plate. The Euler buckling theorem is developed for static compression processes. The static compression process is equivalent to molecular dynamics (MD) simulations with extremely low strain rates. Although this theorem has been widely used in static mechanical processes, it is still unclear whether the Euler buckling theorem is applicable in dynamical compression processes, where the strain rate has important effect on the compression/tension behavior of the system.\cite{JiangJW2012jmps} For instance, it is crucial to apply mechanical strain at a very low strain rate for the study of structure transitions, so that the system has enough time to relax its structure. We thus examine in present work whether the Euler buckling theorem is applicable in the dynamical compression process.

In this letter, we perform MD simulations to examine the applicability of the Euler buckling theorem in the dynamical compression of SLMoS$_{2}$ at different strain rates. It turns out that the Euler buckling theorem is not applicable for longer SLMoS$_{2}$ at higher strain rate, in which the buckling critical strain becomes length independent. However, the Euler buckling theorem will become applicable after restricting it to individual ripples in the buckled SLMoS$_{2}$.


The SLMoS$_{2}$ can be constructed by duplicating a rectangular unit cell of (5.40, 3.12)~{\AA} in the two-dimensional plane. The number of unit cell is $(n_{x}, n_{y})$ in the armchair and zigzag directions. The length of the SLMoS$_{2}$ is $5.40\times n_{x}$, and its width is $3.12\times n_{y}$.  We fix $n_{y}=10$ for all simulations in present work. The free boundary condition is applied in the out-of-plane direction. We apply the fixed boundary condition in armchair direction and periodic boundary condition in zigzag direction. The SLMoS$_{2}$ is compressed unaxially along the armchair direction. The zigzag direction is kept stress free during compression.

All MD simulations in this work are performed using the publicly available simulation code LAMMPS~\cite{PlimptonSJ}, while the OVITO package was used for visualization.~\cite{ovito} The standard Newton equations of motion are integrated in time using the velocity Verlet algorithm with a time step of 1~{fs}. The interaction within MoS$_{2}$ is described by the Stillinger-Weber potential.\cite{JiangJW2013sw} All simulations are performed at 1.0~K low temperature, so that our MD simulations are more comparable with the Euler buckling theorem, which does not consider the temperature effect. The SLMoS$_{2}$ is thermalized using the Nos\'e-Hoover\cite{Nose,Hoover} thermostat for 100~{ps} within the NPT (i.e. the number of particles N, the pressure P and the temperature T of the system are constant) ensemble. After thermalization, the SLMoS$_{2}$ is compressed unaxially along the armchair direction, while the system is allowed to be fully relax in the zigzag direction. NPT ensemble is applied also applied in the compression step.

\begin{figure}[htpb]
  \begin{center}
    \scalebox{1}[1]{\includegraphics[width=8cm]{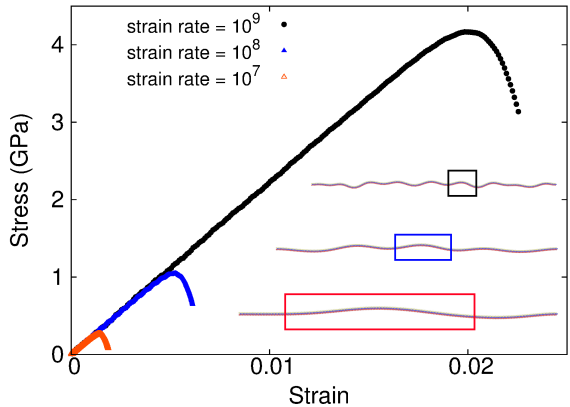}}
  \end{center}
  \caption{(Color online) Strain rate effect on the buckling critical strain of SLMoS$_{2}$. Insets (from top to bottom) illustrate the buckling mode of the SLMoS$_{2}$ at strain rates of $10^{9}$, $10^8$, and $10^7$~{s$^{-1}$}, respectively. The size of a single buckling ripple is enclosed by rectangulars.}
  \label{fig_stress_strain}
\end{figure}

Fig.~\ref{fig_stress_strain} shows the stress-strain relation for the SLMoS$_{2}$ with $n_x=100$, which is compressed at strain rates of $\dot{\epsilon}=$ $10^{9}$, $10^8$, and $10^7$~{s$^{-1}$}, respectively. A value for the thickness is required for the computation of the stress. However thickness is not a well-defined quantity in one-atomic thick layered materials such as SLMoS$_{2}$. Hence, we have assumed the thickness of the SLMoS$_{2}$ to be the space between two neighboring MoS$_{2}$ layers in the three-dimensional bulk MoS$_{2}$. That is the thickness is 6.09~{\AA} for SLMoS$_{2}$. The x-axis is the absolute value for the compression strain. The SLMoS$_{2}$ buckles at the critical strain, at which the stress within the system starts to drop. The critical strain is sensitive to the strain rate, and the buckling critical strain increases sharply with increasing strain rate.

Insets (from top to bottom) of Fig.~\ref{fig_stress_strain} illustrate the buckling mode of the SLMoS$_{2}$, which is compressed with strain rates of $\dot{\epsilon}=$ $10^{9}$, $10^8$, and $10^7$~{s$^{-1}$}, respectively. An individual ripple in the buckling mode is enclosed by rectangular. The length of the ripple decreases quickly with increasing strain rate. Normally, the buckling mode follows the shape of the first bending phonon mode in the system, in which only one ripple occurs after buckling. However, if the system is compressed very fast (i.e. with high strain rate), the buckling mode does not follow the shape of the first bending phonon mode of the SLMoS$_{2}$, and there will be more than one ripple in the buckling SLMoS$_{2}$. In other words, higher-energy bending modes are actuated by the fast compression. This phenomenon has also been observed in the compression of graphene.\cite{Neek-AmalM2010prb}

\begin{figure}[htpb]
  \begin{center}
    \scalebox{1}[1]{\includegraphics[width=8cm]{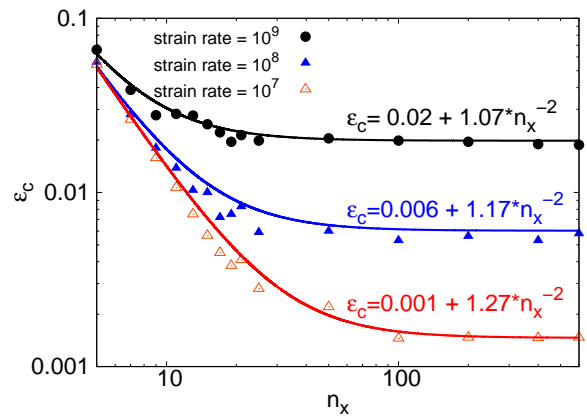}}
  \end{center}
  \caption{(Color online) Length dependence (log-log) of the buckling critical strain for the SLMoS$_{2}$ at strain rates of $10^{9}$, $10^8$, and $10^7$~{s$^{-1}$}, respectively. Simulated data are fitted to functions $\epsilon_c=a+bn_x^{-2}$, which becomes a length-independent constant $a$ in the limit of $n_x \rightarrow +\infty$.}
  \label{fig_critical_strain}
\end{figure}

This strain rate effect can be interpreted in terms of the relaxation time for each bending mode. The first bending modes has the longest relaxation time (or oscillation period), $\tau=2\pi/\omega$, due to its lowest angular frequency $\omega$. It means that the longest response time is needed for the appearance of the first bending mode during the compression of the SLMoS$_{2}$. When the system is compressed very fast, the response time is too short for the occurrence of the first bending mode. Instead, higher-energy bending modes have shorter relaxation, and is able to be actuated by buckling when the SLMoS$_{2}$ is subjected to a fast compression.

Fig.~\ref{fig_critical_strain} shows the buckling critical strain for SLMoS$_{2}$ of different length. The system is compressed at three different strain rates. The simulation data are fitted to the function $\epsilon_c=a+bn_x^{-2}$. The second term $n_x^{-2}$ obeys the Euler buckling theorem, which says the critical strain is an inverse quadratic function of the system length.\cite{TimoshenkoS1987} It means that the Euler buckling theorem is valid for short systems. However, in the limit of $n_x \rightarrow +\infty$, the critical strain becomes a length-independent constant $a=$ 0.0198, 0.0060, and 0.0015 for strain rates of $10^{9}$, $10^8$, and $10^7$~{s$^{-1}$}, respectively. This saturating phenomenon clearly demonstrates that the Euler buckling theorem is not applicable in such dynamical process. For $\dot{\epsilon}=10^7$~{s$^{-1}$}, the critical strain is almost saturate when $n_x>100$. For higher strain rates, the saturation of the critical strain happens at shorter length. That is, the critical strain becomes a constant when $n_x >50$ for $\dot{\epsilon}=10^8$~{s$^{-1}$}, and $n_x > 15$ for $\dot{\epsilon}=10^9$~{s$^{-1}$}. The fitting parameter $a$ becomes closer to zero for lower strain rate, as the dynamical process is more similar as a static process when the strain rate is lower.

\begin{figure}[htpb]
  \begin{center}
    \scalebox{1}[1]{\includegraphics[width=8cm]{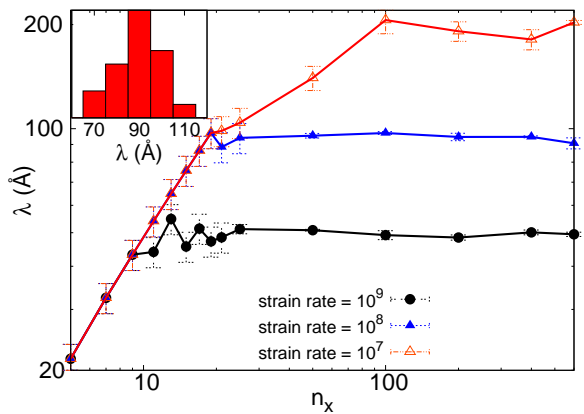}}
  \end{center}
  \caption{(Color online) The length dependence of the average ripple size ($\lambda$) for SLMoS$_{2}$ compressed by strain rates of $10^{9}$, $10^8$, and $10^7$~{s$^{-1}$}, respectively. Inset shows the distribution of the ripple size for a buckling SLMoS$_{2}$ with $n_x=600$, which is compressed at $\dot{\epsilon}=10^8$~{s$^{-1}$}. Twenty good ripple samples have been picked out from the buckled SLMoS$_{2}$ for the production of this histogram plot. The averaged ripple size from the histogram figure is $\lambda=90.7\pm 3.3$~{\AA}. Lines are guide to the eye.}
  \label{fig_ripple_size}
\end{figure}

\begin{figure}[htpb]
  \begin{center}
    \scalebox{1}[1]{\includegraphics[width=8cm]{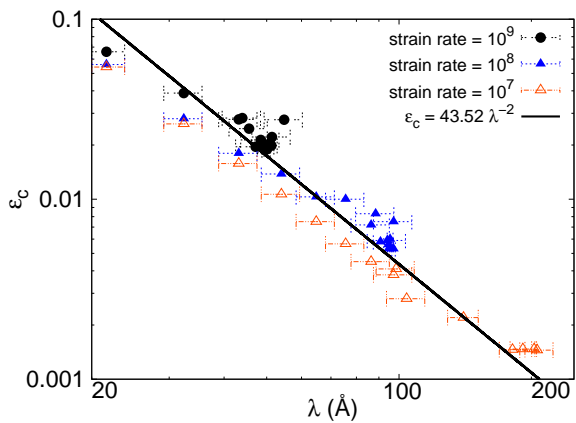}}
  \end{center}
  \caption{(Color online) Buckling critical strain versus buckling ripple size for SLMoS$_{2}$ compressed by strain rates of $10^{9}$, $10^8$, and $10^7$~{s$^{-1}$}, respectively. The solid line is the prediction of the Euler buckling theorem. All simulation data are close to the solid line, which validates the Euler buckling theorem.}
  \label{fig_critical_strain_ripple_size}
\end{figure}

To explore the origin for the inapplicability of the Euler buckling theorem, we examine the ripples in the buckling mode. We first count the number of ripples in the buckling mode. For $\dot{\epsilon}=10^7$~{s$^{-1}$}, there is only one ripple in the buckled SLMoS$_{2}$ with $n_x<100$. For $\dot{\epsilon}=10^8$~{s$^{-1}$}, the buckled SLMoS$_{2}$ has only one ripple if $n_x <50$. For $\dot{\epsilon}=10^9$~{s$^{-1}$}, there is only one ripple in the buckled SLMoS$_{2}$ with $n_x<15$, and more than one ripple is observed for longer systems with $n_x>15$. For instance, the insets of Fig.~\ref{fig_stress_strain} show that there are many ripples in the buckled SLMoS$_{2}$ with $n_x=100$, when this system is compressed at a strain rate of $10^9$~{s$^{-1}$}. These ripples are utilized to get the averaged ripple size. It should be noted that, for short systems with only one ripple in the buckling mode, the length of the SLMoS$_{2}$ will be regarded as the averaged ripple size, and the error is simply chosen as $10\%$ of the length in this situation.

Fig.~\ref{fig_ripple_size} shows the length dependence of the averaged ripple size for SLMoS$_{2}$  compressed at different strain rates. Fig.~\ref{fig_ripple_size} inset shows the distribution of the ripple size for a buckling SLMoS$_{2}$ with $n_x=600$, which is compressed at $\dot{\epsilon}=10^8$~{s$^{-1}$}. Twenty good ripple samples have been picked out from the buckled SLMoS$_{2}$ for the production of this histogram plot. The averaged ripple size from the histogram figure is $\lambda=90.7\pm 3.3$~{\AA}. Fig.~\ref{fig_ripple_size} shows an interesting phenomenon that the averaged ripple size is almost saturated for longer systems with larger $n_x$, where more than one ripple occurs. For higher strain rate, the averaged ripple size becomes saturated at smaller length. This saturation phenomenon is similar as the length dependence of the buckling critical strain shown in Fig.~\ref{fig_critical_strain}, which has demonstrated the inapplicability of the Euler buckling theorem.

Intrigued by the saturation phenomena in both Figs.~\ref{fig_critical_strain} and ~\ref{fig_ripple_size}, we find that the Euler buckling theorem is closely related to the number of ripples in the buckling mode. It is valid in case of only one ripple in the buckling mode. However, the Euler buckling theorem becomes invalid when more than one ripple occurs. The Euler buckling theory says,\cite{TimoshenkoS1987}
\begin{eqnarray}
\epsilon_{c}  =  -\frac{4\pi^{2}D}{C_{11}L^{2}} = -\frac{43.52}{L^{2}},
\label{eq_euler}
\end{eqnarray}
where $L$ is the length of the system. We have used the Stillinger-Weber potential to extract the bending modulus\cite{JiangJW2013bend}  $D=9.61$~{eV} and the in-plane tension stiffness\cite{JiangJW2013sw} $C_{11}=139.5$~{Nm$^{-1}$} for the SLMoS$_{2}$. We note an important fact that only one ripple is assumed in the buckling mode during the derivation of Eq.~(\ref{eq_euler}). However, Fig.~\ref{fig_ripple_size} discussed the suitability of the Euler buckling theorem based on regarding $L$ as the total length of the system. It seems that a more proper way is to treat $L$ as the size of an individual ripple in the buckling mode with many ripples. We thus show the buckling critical strain v.s averaged ripple size in Fig.~\ref{fig_critical_strain_ripple_size}. The prediction of the Euler buckling theorem is also plotted in the figure (black solid line) for comparison. We find that all simulation data (calculated with different strain rates) are closely distributed around the line for the Euler buckling theorem. In other words, the Euler buckling theorem is applicable and is independent of the strain rate, after we have treated $L$ in Eq.~(\ref{eq_euler}) as the averaged ripple size. The merit of using lower strain rate ($10^7$~{s$^{-1}$}) is to extend the examination of the Euler buckling theorem to larger ripple size.

We have performed MD simulations to investigate whether the Euler buckling theorem is applicable for dynamical processes, in which the SLMoS$_{2}$ is compressed at high strain rates. We found that the theorem is not applicable in the presence of more than one ripple in the buckling mode, where the buckling critical strain becomes a length-independent constant. However, we have also showed that the Euler buckling becomes applicable if this theorem is applied to a single ripple in the buckled SLMoS$_{2}$.

\textbf{Acknowledgements} The work is supported by the Recruitment Program of Global Youth Experts of China and the start-up funding from Shanghai University. The author thanks T.-Z. Zhang and X.-M. Guo for insightful discussions.

%
\end{document}